%% file: main.tex
\documentclass[conference]{IEEEtran}
\IEEEoverridecommandlockouts
\usepackage{cite}
\usepackage{amsmath,amssymb,amsfonts}
\usepackage{algorithmic}
\usepackage{graphicx}
\usepackage[capitalise,nameinlink,noabbrev]{cleveref}
\usepackage{textcomp}
\usepackage{booktabs}
\usepackage{url}
\usepackage{multicol}
\usepackage{xcolor}
\def\BibTeX{{\rm B\kern-.05em{\sc i\kern-.025em b}\kern-.08em
    T\kern-.1667em\lower.7ex\hbox{E}\kern-.125emX}}

\usepackage{glossaries}

\newcommand\todo[1]{\textcolor{black}{#1}\PackageWarning{}{#1!}}
\newcommand\changes[1]{\textcolor{black}{#1}\PackageWarning{}{#1!}}

\newif\ifreviewmode

\input{acronyms}

\begin{document}

\title{AraXL: A Physically Scalable, Ultra-Wide RISC-V Vector Processor Design for Fast and Efficient Computation on Long Vectors
\ifreviewmode
\else
\thanks{\textcopyright 2025 IEEE. Personal use of this material is permitted. Permission from IEEE must be
obtained for all other uses, in any current or future media, including
reprinting/republishing this material for advertising or promotional purposes, creating new
collective works, for resale or redistribution to servers or lists, or reuse of any copyrighted
component of this work in other works.}
\thanks{}
\thanks{\textsuperscript{*} The first two authors contributed equally to this work.}
\fi
}

\ifreviewmode
\author{\emph{Hidden for double-blind review purposes.}}
\else
\author{\IEEEauthorblockN{Navaneeth Kunhi Purayil\textsuperscript{*}\textsuperscript{1}, Matteo Perotti\textsuperscript{*}\textsuperscript{1}, Tim Fischer\textsuperscript{1} and Luca Benini\textsuperscript{1,2}}
\IEEEauthorblockA{\textsuperscript{1}\textit{ETH Zürich, Zürich, Switzerland}, \textsuperscript{2}\textit{Università di Bologna, Bologna, Italy}}

\{nkunhi,mperotti,fischeti,lbenini\}@iis.ee.ethz.ch}
\fi

\maketitle

\bstctlcite{IEEE:BSTcontrol}

\begin{abstract}
\input{abstract}
\end{abstract}


\section{Introduction}
\label{sec:intro}
\input{introduction}

\section{Related works}
\label{sec:relworks}
\input{related_works}

\section{Architecture}
\label{sec:arch}
\input{architecture}

\section{Evaluation}
\label{sec:eval}
\input{evaluation}

\section{Conclusions}
\label{sec:conclusions}
\input{conclusions}

\section{Acknowledgments}
\label{sec:ack}
\input{ack}

\bibliographystyle{IEEEtran}
\bibliography{references,ieeetran}

\end{document}

%% file: acronyms.tex
\newacronym{asic}{ASIC}{application-specific integrated circuit}
\newacronym{dft}{DFT}{design for testability}
\newacronym{dz}{DZ}{Microelectronics Design Center at ETH Zurich}
\newacronym{fpga}{FPGA}{field-programmable gate array}
\newacronym{ic}{IC}{integrated circuit}
\newacronym{iis}{IIS}{Integrated Systems Laboratory}
\newacronym{snr}{SNR}{signal-to-noise ratio}
\newacronym{hpc}{HPC}{High Performance Computing}
\newacronym{ppa}{PPA}{Power Performance Area}
\newacronym{cmos}{CMOS}{complementary metal-oxide semiconductor}
\newacronym{simd}{SIMD}{Single Instruction Multiple Data}
\newacronym{sse}{SSE}{Streaming SIMD Extensions}
\newacronym{avx}{AVX}{Advanced Vector Extensions}
\newacronym{sve}{SVE}{Scalable Vector Extension}
\newacronym{rvv}{RVV}{RISC-V Vector Extension}
\newacronym{fpu}{FPU}{Floating Point Unit}
\newacronym{vrf}{VRF}{Vector Register File}
\newacronym{VRF}{VRF}{Vector Register File}
\newacronym{rtl}{RTL}{Register Transfer Level}
\newacronym{xbar}{XBAR}{Crossbar}
\newacronym{cmg}{CMG}{Core Memory Group}
\newacronym{vpp}{VPP}{Vector Parallel Pipeline}
\newacronym{alu}{ALU}{Arithmetic Logic Unit}
\newacronym{fma}{FMA}{Floating Multiply Accumulate}
\newacronym{llc}{LLC}{Last Level Cache}
\newacronym{hbm}{HBM}{High Bandwidth Memory}
\newacronym{vpu}{VPU}{Vector Processing Unit}
\newacronym{vlen}{VLEN}{Vector Length}
\newacronym{dlen}{DLEN}{Datapath Width}
\newacronym{elen}{ELEN}{Element Length}
\newacronym{isa}{ISA}{Instruction Set Architecture}
\newacronym{csr}{CSR}{Control Status Register}
\newacronym{vl}{VL}{Vector Length}
\newacronym{ew}{EW}{Element Width}
\newacronym{sldu}{SLDU}{Slide Unit}
\newacronym{tlp}{TLP}{Thread-Level Parallelism}
\newacronym{masku}{MASKU}{Mask Unit}
\newacronym{vlsu}{VLSU}{Vector Load Store Unit}
\newacronym{vldu}{VLDU}{Vector Load Unit}
\newacronym{vstu}{VSTU}{Vector Store Unit}
\newacronym{axi}{AXI}{Advanced eXtensible Interface}
\newacronym{addrgen}{ADDRGEN}{Address Generator}
\newacronym{sram}{SRAM}{Static Random Access Memory}
\newacronym{lmul}{LMUL}{Length Multiplier}
\newacronym{tt}{TT}{Typical Typical}
\newacronym{ss}{SS}{Slow Slow}
\newacronym{vcd}{VCD}{Value Change Dump}
\newacronym{ml}{ML}{Machine Learning}
\newacronym{riscv}{RISC-V}{Reduced instruction set Computer}
\newacronym{DP}{DP}{double-precision}
\newacronym{dnn}{DNN}{Deep Neural Networks}
\newacronym{PPA}{PPA}{power, performance, and area}
\newacronym{GLSU}{GLSU}{Global Load Store unit}
\newacronym{RINGI}{RINGI}{Ring Interface}
\newacronym{REQI}{REQI}{Request Interface}
\newacronym{A2A}{A2A}{all-to-all}
\newacronym{dlp}{DLP}{Data-Level Parallelism}

%% file: abstract.tex
The ever-growing scale of data parallelism in today's HPC and ML applications presents a big challenge for computing architectures' energy efficiency and performance. Vector processors address the scale-up challenge by decoupling Vector Register File (VRF) and datapath widths, allowing the VRF to host long vectors and increase register-stored data reuse while reducing the relative cost of instruction fetch and decode. However, even the largest vector processor designs today struggle to scale to more than 8 vector lanes with double-precision Floating Point Units (FPUs) and 256 64-bit elements per vector register. This limitation is induced by difficulties in the physical implementation, which becomes wire-dominated and inefficient.

In this work, we present AraXL, a modular and scalable 64-bit RISC-V V vector architecture targeting long-vector applications for HPC and ML. AraXL addresses the physical scalability challenges of state-of-the-art vector processors with a distributed and hierarchical interconnect, supporting up to 64 parallel vector lanes and reaching the maximum Vector Register File size of 64 Kibit/vreg permitted by the RISC-V V 1.0 ISA specification. Implemented in a 22-nm technology node, our 64-lane AraXL achieves a performance peak of \todo{146 GFLOPs} on computation-intensive HPC/ML kernels ($>$99\% FPU utilization) and energy efficiency of \todo{40.1} GFLOPs/W (\todo{1.15} GHz, TT, 0.8V), with only \todo{3.8$\times$} the area of a 16-lane instance.

%% file: introduction.tex
The amount of data and the computational needs of today's applications have skyrocketed, with no signs of slowing down. This unprecedented growth requires innovative solutions in hardware and software, as technology scaling alone no longer provides a reliable way of boosting chips' performance. Moreover, simply chasing performance through frequency improvements is no longer viable due to the power wall \cite{7013055}.
As a result, prioritizing energy efficiency over sheer performance has become imperative in modern hardware designs, even in the \gls{hpc} domain \cite{efficientHPC}.

To address this challenge, one of the most efficient solutions is leveraging applications' \gls{dlp} to encode multiple operations in a single instruction, amortizing its fetch/decode cost, as
many crucial \gls{hpc} and \gls{ml} applications exhibit high degrees of parallelism and often require computation on extremely long vectors. 

\gls{simd} architectures are able to exploit long vector lengths by simultaneously computing multiple vector elements with a single instruction. However, these architectures are limited by their \gls{vrf} width (i.e., the number of vector elements that can be buffered in the architecture), which is \changes{usually} the datapath width, hindering the data reuse and the amortization of instruction fetch and decode.

On the other hand, Cray-inspired \cite{Russell1978} vector processor architectures feature \glspl{vrf} whose size is decoupled from the datapath width. The vector length can be programmed at runtime and is upper-bounded by the physical size of each vector register, which is a design parameter.
Supporting large vector lengths not only minimizes the energy spent in fetching, decoding, and issuing instructions but also has critical performance benefits. Longer vectors lower pressure on the data memory due to higher data reuse close to the \glspl{fpu}. Further, they exhibit a higher tolerance for stalls and memory latency, resulting in improved performance for both dense \cite{ramirez_risc-v_2020} and sparse workloads \cite{gupta_challenges_2023, vizcaino_short_2023, gomez_efficiently_2021, gomez_hpcg_2023}. 

Due to these reasons, vector processor architectures are gaining traction.  For instance, Arm developed the \gls{sve} (2016) and \gls{sve}2 (2019) scalable vector \gls{isa} extensions, used in the Arm Neoverse V2 and the Fujitsu A64FX cores to power the AWS GRAVITON4 \cite{graviton3} and supercomputer FUGAKU \cite{okazaki2020supercomputer}, respectively.
The open-source RISC-V \gls{isa} has also recently ratified its vector extension V 1.0, with a plethora of novel architectures designed by universities \cite{Platzer2021, Minervini2022, ara2perotti24} and companies \cite{SiFiveX280, SiFiveP870, AndesNX27V}. RISC-V V, more than Arm \gls{sve}, highlights the importance of long vectors, allowing a maximum vector length of 64 Kibit per vector register. However, as of today, no RISC-V vector processor architecture has ever implemented such large \gls{vrf}.

Scaling up the \gls{vrf} and the number of parallel \glspl{fpu} of a vector processor architecture presents numerous challenges. Vitruvius+ \cite{Minervini2022} and Ara2 \cite{ara2perotti24} are the largest RISC-V V vector processor architectures available and feature up to 8 and 16 parallel lanes with one double-precision \gls{fpu} per lane, respectively, and a large \glspl{vrf} with up to 16 Kibit of vector length. Their \gls{vrf} is split among the vector lanes to improve data locality and limit the interconnect complexity. Despite being a modular lane-based architecture, Ara2 showed that scaling to more than 8 lanes is challenging due to numerous all-to-all interconnects that enable data movements between the memory and the lanes (\gls{vrf}) and among the lanes.

In this work, we present AraXL, which leverages long vectors in \gls{hpc} and \gls{ml} applications to tolerate memory latency and therefore ease the physical implementation. AraXL solves the scalability challenge of the all-to-all interconnects with a dedicated hierarchical and pipelined interconnect, allowing it to scale up to 64 double-precision floating-point-capable vector lanes.

The key contributions of this paper are:
\begin{itemize}
    \item AraXL, the first RISC-V vector processor architecture able to scale up to 64 lanes supporting vectors of up to 8192 \gls{DP}-elements. AraXL tolerates memory latency and overcomes today's vector processor scalability limitations, achieving the longest vector length permitted by the RISC-V V ISA specifications.
    \item AraXL's physical implementation in a 22-nm technology node and a study of its \gls{PPA} metrics. AraXL's modular architecture scales up from 2 to 64 lanes with almost perfect area scaling (2$\times$ when doubling the lane count and \gls{vrf} size). The maximum frequency is never lower than 1.15 GHz with an energy efficiency of 40 GFLOPs/W (TT, 0.8V, 25C).
    \item An evaluation of AraXL's performance and tolerance to memory latency on multiple compute- and memory-intensive kernels, showing almost perfect performance scaling under weak scaling conditions. AraXL with 64 lanes can reach up to 99\% utilization on the \texttt{fmatmul} matrix multiplication kernel.
\end{itemize}

%% file: related_works.tex
\begin{figure}
    \centering
    \includegraphics[width=0.7\linewidth]{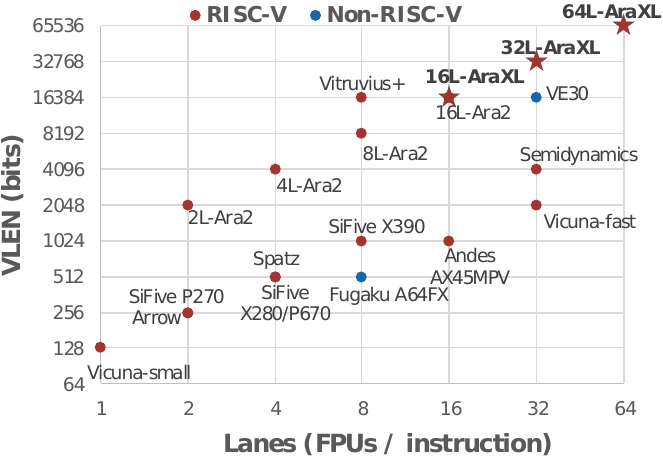}
    \caption{Vector processors grouped by vector register bit-width (VLEN) and number of \gls{fpu} used by a vector instruction.}
    \label{fig:specs}
\end{figure}

Significant progress has been made in the development of scalable and energy-efficient vector processors, particularly following the introduction of the RISC-V V vector extension. \Cref{fig:specs} summarizes the most notable ones, shown by their vector register length and number of processing units (\glspl{fpu}). 

Many of these processors are designed for applications that do not exhibit extremely high vector lengths and typically feature a limited \gls{vrf} and fewer \glspl{fpu}. These designs often focus on smaller workloads and leverage multicore configurations to exploit \gls{tlp} or other dimensions of parallelization. Examples of such processors include most of the SiFive vector architectures \cite{X390, SiFiveX280, SiFiveP270}, Spatz \cite{cavalcante2023spatz}, small instances of Ara2/Vicuna \cite{ara2perotti24, Platzer2021}, and Arrow \cite{assir2021arrow}. These architectures are not primarily aimed at \gls{hpc} applications, where long vectors are exposed, and high \gls{fpu} counts become crucial.

Vector processors targeting the \gls{hpc} domain also exist and typically feature a higher number of \glspl{fpu}. For instance, the Fujitsu A64FX is made of four \glspl{cmg}, each composed of 12 computing cores. Each core has 2 \glspl{fpu} delivering a total of 32 double-precision operations per cycle \cite{okazaki2020supercomputer}. However, A64FX's cores implement Arm \gls{sve}, which limits the vector register length to 2048 bits, preventing aggressive exploitation of the benefits of long vectors. 
Some RISC-V vector processors with multiple \glspl{fpu} also have limited register file capacity: the configurable vector units from \changes{Andes AX45MPV \cite{Andes} and }Semidynamics \cite{Semidynamics} feature \glspl{vrf} up to 16 and 32 \glspl{fpu}, respectively, but with VLEN (bit-width of a vector register) limited to 1024 and 4096 bit.

Vector processors targeting long vectors with higher \gls{fpu} counts exist as well. Notable examples include Vitruvius+, the 8- and 16-lane Ara2 instances, and NEC's TSUBASA Aurora.

Vitruvius+, part of the European Processor Initiative, is tailored for \gls{hpc} applications that expose long vectors, supporting 8 lanes and a VLEN of 16 Kibits \cite{Minervini2022}.
Ara2 is another RISC-V V processor with a VLEN of up to 16 Kibits and 16 \glspl{fpu}, but scaling Ara2 microarchitecture beyond 8 lanes is challenging due to the complexity of all-to-all interconnects that allow data movements among the lanes (in the load-store, mask, and slide units)  \cite{ara2perotti24}.

Today's largest-scale vector processor is not a RISC-V one, though. NEC’s TSUBASA Aurora explicitly targets extremely long vectors with its multi-core VE30 vector engines \cite{VE30}. A single core features 32 lanes, each with 3 FMAs, 2 ALUs, and a Complex/Store pipeline, totaling 6 execution pipelines connected to a multi-ported \gls{vrf} with a \changes{\acrshort{vlen}} of 16 Kibits. However, despite its theoretical peak performance, its ability to buffer elements in the register file is limited compared to the RISC-V V specifications. Moreover, the microarchitecture of the VE30 is proprietary, and effective performance and power efficiency have not been benchmarked in the open literature, making it unclear to what extent their performance and efficiency meet theoretical peaks in practical workloads.

Our architecture, AraXL, is designed to tackle the scalability challenges observed in current high-VLEN RISC-V vector processors by means of implementation-friendly and modular interconnects. This allows AraXL to feature 64 \glspl{fpu} and the highest VLEN allowed by RISC-V V (64Kibit per register), maximizing the latency tolerance and power benefits from \gls{hpc} and \gls{ml} long-vector applications. \changes{In fact, higher \acrshort{vlen}s up to 64 Kibit can increase the \acrshort{hpc}-workload performance \cite{gomez_hpcg_2023} and allow leveraging context windows as large as 128k elements in Llama3 \cite{llama3} in the \acrshort{ml} domain.}

%% file: architecture.tex
\subsection{Architecture overview}
AraXL is a decoupled vector processor architecture composed of a CVA6 scalar core \cite{Zaruba2019} and multiple physical clusters that act as a single RISC-V V accelerator. 

\begin{figure}[ht]
    \centering
    \includegraphics[width=1\linewidth]{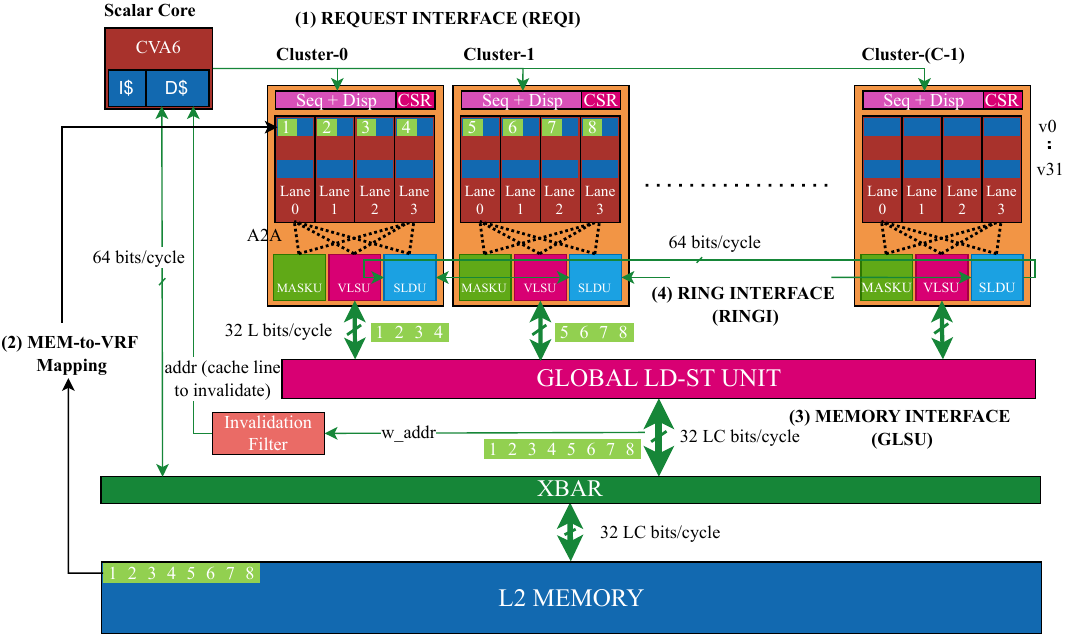}
    \caption{Schematic of the cluster-based AraXL architecture and its interfaces.}
    \label{fig:AraXLarch}
\end{figure}

As depicted in \Cref{fig:AraXLarch}, AraXL's hierarchical architecture is based on vector clusters, each composed of an enhanced instance of the open-source Ara2 \cite{ara2perotti24} equipped with its own dispatcher, sequencer, \gls{A2A}-interconnected units (\gls{masku}, \gls{sldu} and \gls{vlsu}), and lanes, which feature the processing units and the \gls{VRF} chunks. 
We choose the 4-lane configuration as a building block for the vector cluster, as it features the highest energy efficiency \cite{ara2perotti24} among all the configurations, and we modify it by streamlining its internal interconnects.

We design three scalable interfaces - the \gls{REQI}, \gls{GLSU} interface, and \gls{RINGI} - to connect the clusters to the CVA6 scalar core, the L2 memory, and the neighbor clusters for permutation operations, respectively.

Conceptually, AraXL is a set of vector processor clusters synchronized through the \gls{REQI}, operating on vector elements mapped from memory to the internal sparse \gls{VRF} by the \gls{GLSU}, and using the \gls{RINGI} to move data among different clusters. 
To achieve a scalable architecture, we prioritize relaxing the timing of all top-level interconnects over their latency, which is not critical in long-vector applications. This also means that all the possibly-critical paths through the interfaces can be cut with a parametric number of registers, as shown in \Cref{sec:eval}.

We design AraXL to target maximum performance on the most common subset of \gls{rvv} instructions, namely unit-stride memory accesses, slide-by-1 operations, reductions, and basic mask operations. These instructions are the most prevalent in regular data parallel workloads in the \gls{hpc} and \gls{ml} domain. 
Strided/Indexed memory operations, variable slides, and other irregular \gls{rvv} operations are supported, albeit at lower throughput.

In the following sections, we discuss AraXL's \changes{architecture}.

\subsection{Architecture details}

\subsubsection{\gls{REQI}}
CVA6 broadcasts the vector instructions fetched by its L1 i-cache to all clusters through the \gls{REQI}. Then, the clusters decode and execute every instruction in sync. 

Once the request is accepted by clusters, cluster-0 sends the answer back to CVA6, signaling possible exceptions or forwarding scalar results to be written to scalar registers. 

\subsubsection{Memory to VRF byte mapping}
Ara2 implements an element-wise mapping of memory bytes to lanes, such that element-\textit{i} always maps to lane-\textit{$i (mod) L$}, regardless of the \gls{ew}. This ensures that common mixed-width operations, e.g., when accumulations are done in higher precision, do not require scrambling of bytes among lanes \cite{ara2perotti24}. 
AraXL naturally extends Ara2's byte mapping such that element-\textit{i} maps to cluster-\textit{$i/L (mod) C$}, lane-\textit{$i (mod) L$}. 
\Cref{fig:AraXLarch} shows AraXL's memory-\gls{VRF} byte mapping, which happens in two stages: 1) from the memory to the clusters through the \gls{GLSU}, and 2) from each cluster to its lanes via the local \gls{vlsu} within the cluster.

\subsubsection{\gls{GLSU}}
The \gls{GLSU} receives requests from the local \glspl{vlsu} and generates a wider \gls{axi} request to the memory.
In Ara2, the byte mapping interconnect of the \gls{vlsu} showed limited scalability since $8L$ bytes coming from memory are \gls{A2A} interconnected to each $8L$ byte of every lane to support unaligned load-store AXI requests, resulting in quadratic complexity of $L^2$ for the \gls{vlsu} \cite{ara2perotti24}.

To achieve a scalable \gls{GLSU}, AraXL implements the shuffling and aligning logic in a multi-level pipeline to move the memory bytes to the correct clusters in multiple cycles. By integrating this pipelined logic into the interconnect, we trade off latency with higher scalability, which is enabled by the latency tolerance of our target applications.

\begin{figure}[ht]
    \centering
    \includegraphics[width=0.8\linewidth]{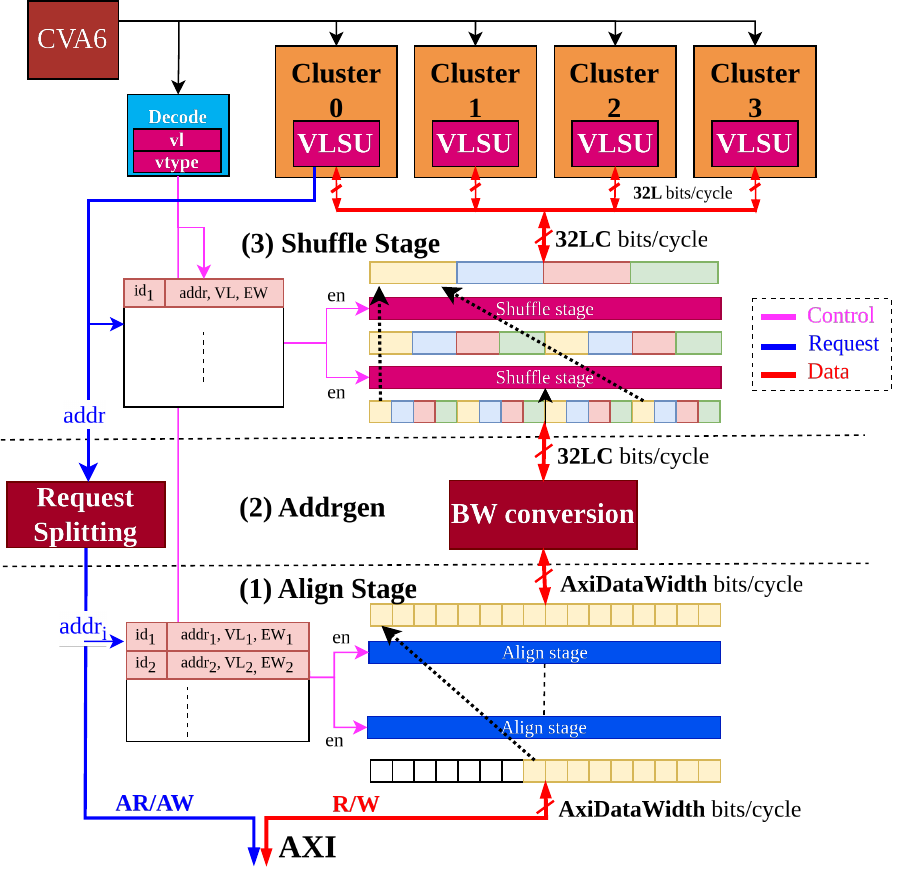}
    \caption{Schematic of the Global Load-Store Unit (GLSU).}
    \label{fig:glsu}
\end{figure}

\Cref{fig:glsu} shows the 3-stage architecture of the \gls{GLSU}. The \emph{Align} stage aligns the misaligned data to the memory bus with multiple power-of-2 shifts. The \emph{Addrgen} stage handles request splitting and bandwidth conversion. The \emph{Shuffle} stage shuffles the aligned data to different clusters based on the \gls{ew} configuration. 
Each level of the \emph{Align} and \emph{Shuffle} stages is guarded by registers and receives control signals based on the address, vector length, and element width, which are tracked in the shuffle and align tables. 

As a consequence, the \glspl{vlsu} local to the clusters only need to shuffle the data bytes to the lanes since aligning has already been done by the \gls{GLSU}. In contrast, the original \gls{vlsu} of Ara2 aligns and shuffles the memory bytes to the lane's \glspl{VRF} in a single cycle, leading to scalability challenges.

\subsubsection{\gls{RINGI}}
The original Ara2 design uses a lumped \gls{sldu} for permutation operations between lanes during, for example, vector slides and reductions.

In AraXL, we implement a ring interconnect to move data among clusters and extend each cluster's \acrshort{sldu} to utilize the data from the ring whenever necessary. 
We choose the ring interconnect since most \gls{hpc} and \gls{ml} workloads utilize slide-by-1 operations requiring only data movement between adjacent clusters. Furthermore, the ring interconnect is easily scalable to a large number of vector clusters.

\begin{figure}[ht]
    \centering
    \includegraphics[width=1\linewidth]{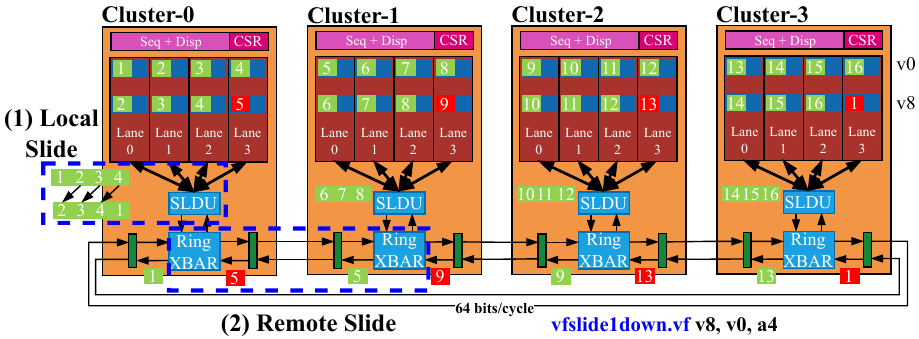}
    \caption{Schematic of the Ring Interface (RINGI).}
    \label{fig:ringi}
\end{figure}

To maximize the performance of the two most common operations (slide1up, slide1down), each cluster supports two output data busses with a bandwidth of 64 bits/cycle, targeting the previous and the next cluster, together with the two incoming ones.
Slides larger than 1 are implemented using multiple 64-bit data transfers or bypasses on the ring to the correct destination lane. A schematic of the \texttt{vfslide1down} operation using the ring is shown in figure \ref{fig:ringi}.

The original Ara2 implements three stages for reductions: intra-lane, inter-lane, and a \gls{simd} stage. In AraXL, we add an inter-cluster stage to reduce the values local to each cluster after the inter-lane step, for which we use the ring interconnect. 
This reduction is done in a log-tree fashion and utilizes multiple hops for later reduction stages. To ensure timing is not affected, AraXL \changes{instantiates} a parametric number of registers in the ring interconnect between \glspl{sldu} of adjacent clusters.

\subsubsection{\gls{masku}}
The original Ara2's \gls{masku} contributed to the scalability issues due to its \gls{A2A} nature at the bit-level to distribute mask bits across different lanes. 
This would be even more problematic in AraXL, where a 64-lane architecture would require distributing each bit of a 64-bit packet to a different lane.

To prevent this, we add a new dedicated \gls{VRF} byte encoding to keep the mask vector bits already in the corresponding lane.

The addition of a new byte layout requires supporting reshuffling conversions between this format and the other supported byte encodings. In AraXL, this is done by the \gls{sldu} through the \gls{RINGI} to move bits across clusters. 

As noted for Ara2, reshuffling is a slow operation, and software should not use the same register to sequentially hold mask and non-mask vector elements to avoid unnecessary byte layout modifications.

%% file: evaluation.tex
\subsection{Evaluation setup}
\begin{figure}
    \centering
    \includegraphics[width=1\linewidth]{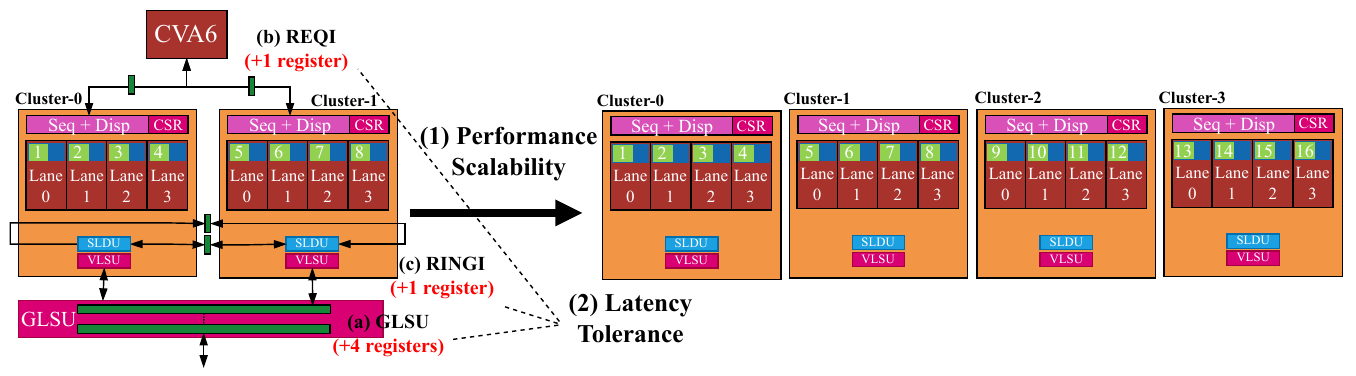}
    \caption{\emph{Performance scalability} and \emph{Latency tolerance} experiment setups.}
    \label{fig:eval_setup}
\end{figure}

We implement AraXL in SystemVerilog and characterize its performance with configurations up to 64 lanes using cycle-accurate simulations with \textsc{Questasim-2021.2}. To benchmark AraXL, we use a selection of common \gls{hpc}/\gls{ml} kernels whose instructions include unit-stride load-stores, slide-by-1 (\texttt{fconv2d}, \texttt{jacobi2d}), reduction (\texttt{fdotproduct}, \texttt{softmax}) and basic mask operations (\texttt{exp}) (\Cref{tab:eval_config}). 
We evaluate AraXL's performance with two metrics: \emph{performance scalability}, with a comparison against the original Ara2, and \emph{latency tolerance} (\Cref{fig:eval_setup}).

Finally, we synthesize and place-and-route AraXL in 22-nm technology with \textsc{Synopsys DC} and \textsc{IC Compiler 2 2022.03} and discuss its \gls{PPA} metrics and scalability for up to 64 lane configurations. We extract the power consumption of the post-layout netlist with power simulations using \textsc{PrimeTime 2022.03} in the typical conditions (0.8V, TT, 25C).

\begin{table}
    \centering    
     \begin{tabular}{rlll}
     \toprule
     Benchmarks & Problem size* & LMUL & Max Perf\\
    & [DP Elements] & &  [DP-FLOP/cycle] \\
     \midrule
     \textbf{fmatmul} & A=64$\times$256 B=256$\times$N & 1,2,4 & 2$\times$LC\\
     \midrule
     \textbf{fconv2d} & A=256$\times$N f=7$\times$7& 2 & 2$\times$LC \\
     \midrule
     \textbf{jacobi2d} & A=256$\times$N & 4 & LC \\
     \midrule
     \textbf{fdotproduct} & A=B=N & 8 & LC\\
     \midrule
     \textbf{exp} & A=N & 1 & 28/21$\times$LC\\
     \midrule
     \textbf{softmax} & A=64$\times$N & 1 & 32/25$\times$LC \\
     \bottomrule
     \multicolumn{4}{l}{* N $= nLC$ (for $L$ lanes per cluster, $C$ clusters, and $n = 16 \times$ LMUL)}\\
     \multicolumn{4}{l}{* \changes{Assuming an L2 memory size of at least 16 MiB to fit the benchmarks}}\\
    \end{tabular}
    \caption{Benchmark parameters used for evaluations.\\}
    \label{tab:eval_config}
\end{table}

\subsection{Performance Scalability}
\label{subsection:perf_scal}
\begin{figure}
    \centering
    \includegraphics[width=1\linewidth]{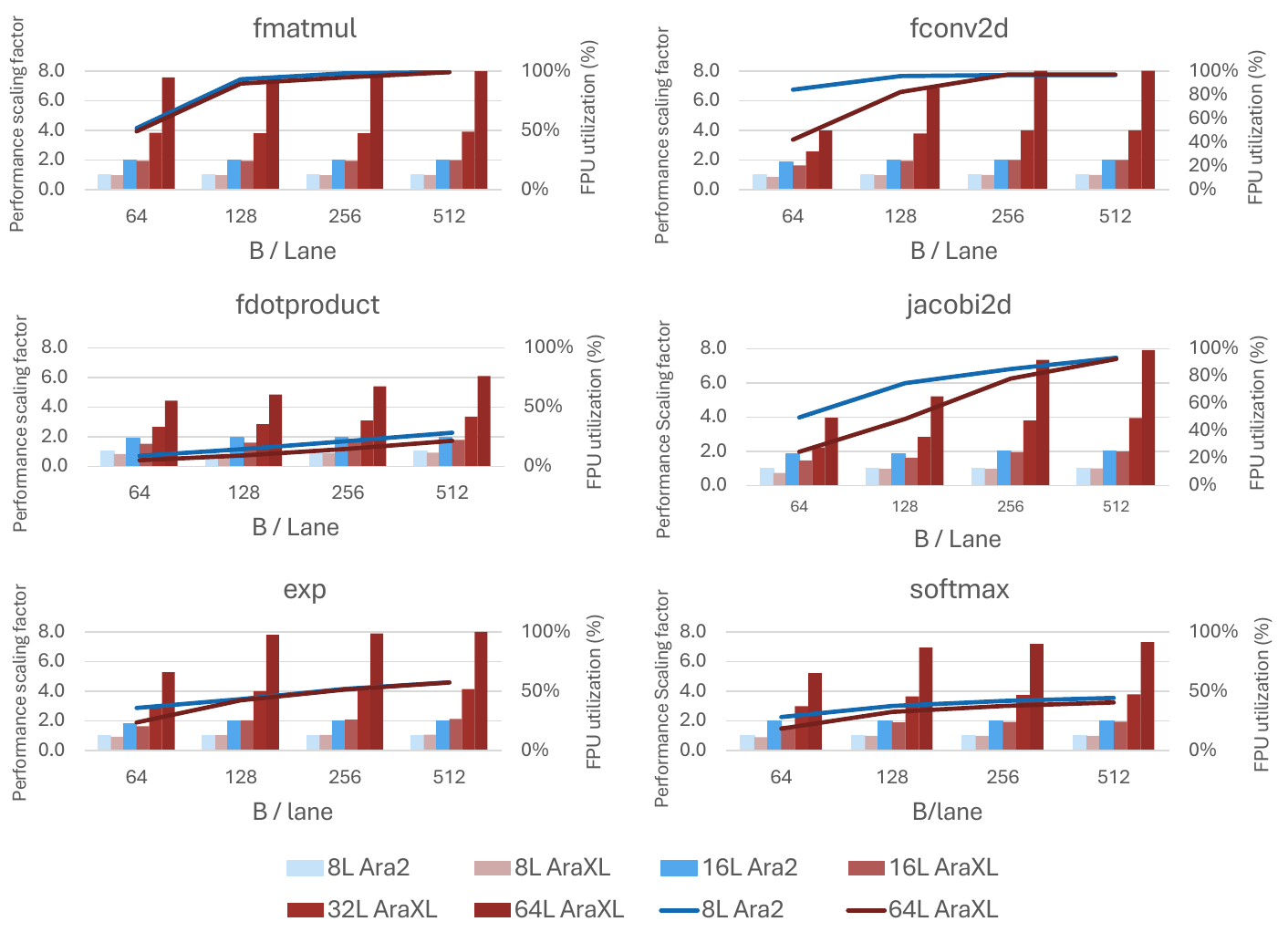}
    \caption{AraXL's performance scalability. The bars (left Y-axis) are normalized on the original 8-lane Ara2's performance. The lines (right Y-axis) represent absolute FPU utilization.}
    \label{fig:perf_scal}
\end{figure}

We characterize AraXL's performance scalability under weak scaling conditions by simulating our kernels on larger AraXL configurations at proportionally larger problem sizes, ensuring that each lane always operates on the same number of bytes. 
\Cref{fig:perf_scal} reports the measured performance values normalized to the original 8-lane Ara2's performance up to 512 B/lane (bar plot, left Y-axis). 
The \gls{fpu} utilization, measured as the percentage of runtime in which the \gls{fpu} is producing valid results, is also reported for the 64-lane AraXL and 8-lane Ara2 (lines, right Y-axis).

In the medium vector length regime (64 B/lane), both the original Ara2 design and AraXL show lower \gls{fpu} utilization since they cannot hide the latency determined by the setup time of the vector instructions and the latency of scalar loads-stores through the data-cache. This effect is worse in AraXL since the newly designed interfaces increase the vector instruction setup time.

In the long vector regime (from 128 B/lane), which is the explicit target of AraXL, the vector pipeline is busy enough to hide the scalar setup time and interface latencies, leading to high \gls{fpu} utilization on the computationally intensive kernels. As can be seen from the figure, \texttt{fmatmul} and \texttt{fconv2d} achieve up to 99\% and 97\% utilization, respectively, and linear performance scaling from 8 to 64 lanes.
This trend is similar for the other two compute-bound \texttt{exp} and \texttt{jacobi2d} kernels.
On the other hand, \texttt{softmax} and the memory-bound \texttt{fdotproduct} kernels use reduction operations, which incurs a noticeable \gls{fpu} utilization drop even for longer vectors, with performance scaling factors of 7.3$\times$ and 6.1$\times$ on a 64-lane AraXL instance, respectively. 
This slight trend degradation is caused by the non-ideal inter-lane log-tree reduction steps funneled through the ring interconnect.
Since the inter-cluster and inter-lane reduction latencies depend on the architecture's configuration and not on the problem size, an even larger vector length mitigates these non-idealities. For example, AraXL can achieve a close-to-linear performance scaling of 7.6$\times$ with a 16384 B/lane vector dot product, strip-mined over 16 loop iterations, as the time spent to partially reduce the vector elements locally to each lane (intra-lane phase) amply dominates the total reduction time, amortizing the non-ideal inter-lane and inter-cluster steps.

Overall, we conclude that AraXL achieves linear performance scaling from 8 to 64 lanes when processing long-vector workloads for all the benchmarks, with high \gls{fpu} utilization, especially for the crucial \texttt{fmatmul} and \texttt{fconv2d} kernels.

\subsection{Latency tolerance}
\label{subsection:lat_tolerance}
\begin{figure}
    \centering
    \includegraphics[width=1.0\linewidth]{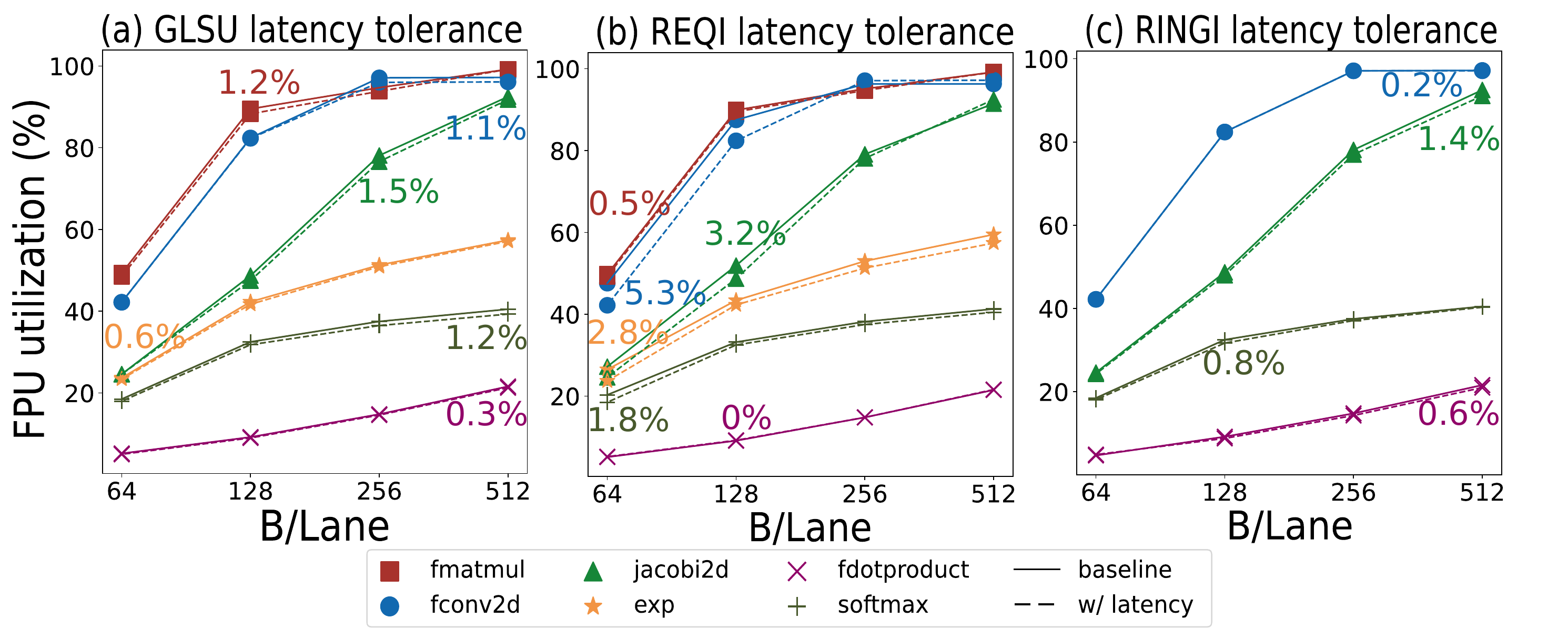}
    \caption{Performance impact of additional latency on the (a) Memory, (b) Request, and (c) Ring interfaces over the 64-lane AraXL baseline. For each kernel, the max. FPU utilization drop is reported.}
    \label{fig:lat_tolerance}
\end{figure}

We also evaluate the impact of the additional latency caused by the insertion of sequential cuts into the cluster interfaces as depicted in \Cref{fig:eval_setup}.
\Cref{fig:lat_tolerance} shows the latency tolerance of AraXL in terms of \gls{fpu} utilization degradation with the addition of 4, 1, and 1 register cuts to the \gls{GLSU}, \gls{REQI}, and \gls{RINGI} interfaces, respectively. 

\paragraph{\gls{GLSU} interface}
The register additions along the \gls{GLSU} interface increase the request-response latency by 8 cycles. As shown in \Cref{fig:lat_tolerance} (a), the maximum utilization drop in the long-vector regime is a mere 1.5\%. Furthermore, longer vectors face virtually no performance drop.

\paragraph{\gls{REQI}}
Adding a register on the \gls{REQI} implies that the vector instruction is acknowledged back to CVA6 2 cycles later, delaying the issue of the next instruction.
From \Cref{fig:lat_tolerance} (b), we see a maximum utilization drop of 5\% for \texttt{fconv2d} and 3\% for \texttt{jacobi2d} at 128 B/lane.
However, this can be completely amortized at 512 B/lane for both kernels.

\paragraph{\gls{RINGI}}
The added registers increase the 1-hop latency between clusters by 1 cycle, which affects slide and reduction operations. 
However, \Cref{fig:lat_tolerance} (c) shows that, for long vectors, we only see up to 1.4\% drop in utilization.

Overall, AraXL exhibits high latency tolerance on all three interfaces - \gls{GLSU}, \gls{REQI} and \gls{RINGI} - achieving less than 2\% utilization drop in the long-vector regime.

\subsection{Physical Implementation}
\begin{figure}
    \centering
    \includegraphics[width=0.8\linewidth]{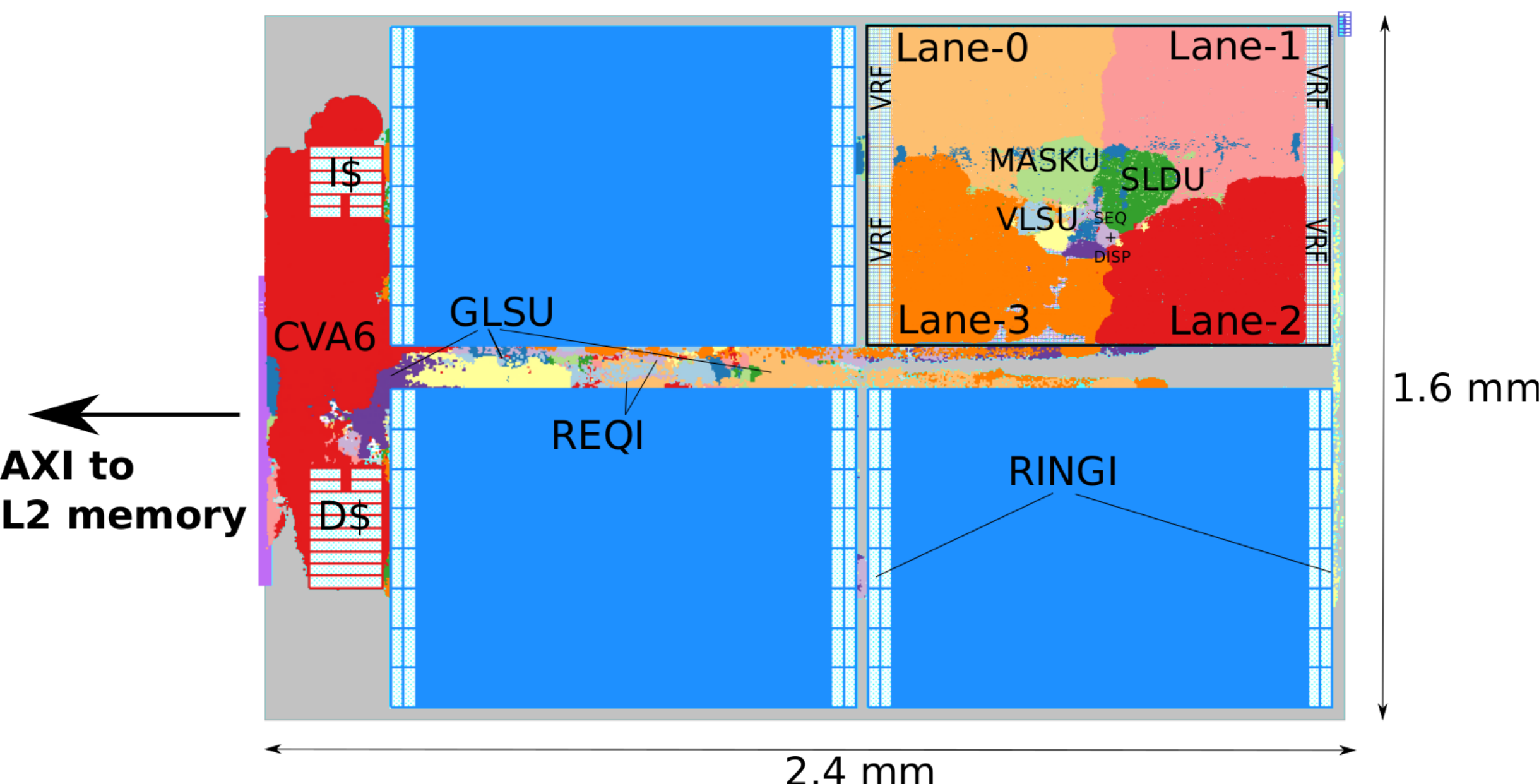}
    \caption{16-lane AraXL floorplan. Focus on the modified 4-lane Ara2 used as a cluster, its \gls{A2A} connected units, CVA6, and the top-level interfaces.}
    \label{fig:16laneFP}
\end{figure}

We perform a hierarchical physical implementation of AraXL in a 22-nm technology with configurations of 16, 32, and 64 lanes and evaluate its \gls{ppa} metrics and scalability. We show the annotated 16-lane AraXL floorplan in \Cref{fig:16laneFP}.

\begin{figure}
    \centering
    \includegraphics[width=1\linewidth]{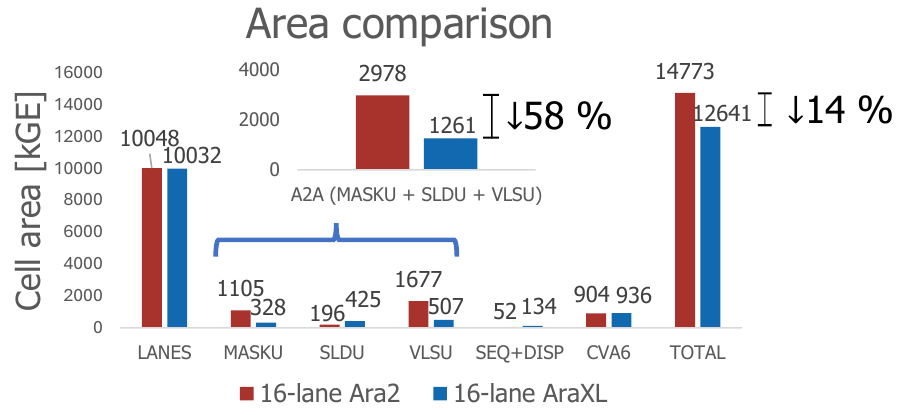}
    \caption{Area breakdown of 16-lane AraXL and Ara2. For a fair comparison, AraXL's VLSU, SLDU, and SEQ+DISP also include the area of the top-level GLSU, RINGI, and REQI interfaces.}
    \label{fig:16lane_comparison}
\end{figure}

\paragraph{Area and Timing}
In \Cref{fig:16lane_comparison}, we compare the area of a 16-lane AraXL against the original 16-lane Ara2 architecture. 
AraXL achieves a significant area improvement of 14\% from the redesign of the \gls{A2A} units - \gls{masku}, \gls{sldu}, and \gls{vlsu} - which limited the scalability of Ara2 beyond 8 lanes.
AraXL also reaches a higher maximum frequency of 1.4 GHz, an improvement from 1.08 GHz in typical conditions (0.8V, TT, 25C). This is enabled by the memory latency tolerance of AraXL shown in \Cref{subsection:lat_tolerance}, which allows removing Ara2's critical path from the align/shuffle complexity of the \gls{A2A} connected \gls{vlsu} and \gls{masku}.

\Cref{tab:AraXL_area_scaling_comparison} shows the area scaling trends of AraXL from 16 to 64 lanes. We see that both 32- and 64-lane AraXL achieve linear scaling w.r.t 16-lane AraXL thanks to the cost-effective interconnect design. The \gls{GLSU}, \gls{RINGI}, and \gls{REQI} account for only 3\% of the total area. 

AraXL reaches 1.4 GHz up to \todo{32 lanes, with a frequency degradation for the 64-lane design \todo{(1.15 GHz) due to floorplan inefficiencies that result in routing congestion hotspots}.}

\begin{table}
    \centering
    \begin{tabular}{rrrr}
         \toprule
          Cell area [kGE] ($\times$*) & 16L-AraXL & 32L-AraXL & 64L-AraXL \\
          \midrule
         Clusters & 11354 (1.0$\times$) & 22708 (2.0$\times$) & 45415 (2.0$\times$) \\
         CVA6 & 936 (1.0$\times$) & 901 (1.0$\times$) & 931 (1.0$\times$) \\
         GLSU & 291 (1.0$\times$) & 618 (2.1$\times$) & 1385 (2.2$\times$) \\
         RINGI & 25 (1.0$\times$) & 44 (1.8$\times$) & 76 (1.7$\times$) \\
         REQI & 34 (1.0$\times$) & 81 (2.4$\times$) & 144 (1.8$\times$) \\
         \midrule
         \textbf{TOTAL} & 12641 (1.0$\times$) & 24352 (1.9$\times$) & 47950 (2.0$\times$) \\
         \bottomrule
    \multicolumn{4}{l}{* Scaling factor normalized to half the number of lanes}\\
    \end{tabular}
    \caption{AraXL area breakdown and scaling characterization.}
    \label{tab:AraXL_area_scaling_comparison}
\end{table}

\paragraph{Efficiency comparison}
We calculate the energy efficiency metrics for 16-, 32-, and 64-lane AraXL simulating \texttt{fmatmul} in the long-vector regime (512 B/lane) in the typical conditions (0.8V, TT, 25C). 
As shown in \Cref{tab:efficiency}, AraXL achieves an energy efficiency of \todo{40.4 GFLOPS/W} and an area efficiency of \todo{17.8 GFLOPS/\(mm^2\)} showing significant improvements w.r.t 16-lane Ara2.

\begin{table}
    \centering
    \begin{tabular}{rlllll}
         \toprule
          & L & Freq.* & Max. Perf. & Energy Eff. & Area Eff. \\
          & & [$\mathrm{GHz}$] & [$\mathrm{GFLOPs}$] & [$\mathrm{\frac{GFLOPs}{W}}$] & [$\mathrm{\frac{GFLOPs}{mm^2}}$] \\
          \midrule
         Vitruvius+  & 8 & 1.40 & 22.4 & 47.3** & 17.23** \\
         Ara2 & 16 & 1.08 & 34.2 & 30.3 & 11.6 \\
         \textbf{AraXL}  & 16 & 1.40 & 44.3 & 39.6 & 17.4 \\
         \textbf{AraXL}  & 32 & 1.40 & 87.2 & 40.4 & 17.8 \\
         \textbf{AraXL}  & 64 & \todo{1.15} & \todo{146.0} &  \todo{40.1} &  \todo{15.1} \\
         \bottomrule
    \multicolumn{6}{l}{* Typical corner max freq. ** Scalar core and caches not included.}\\
    \end{tabular}
    \caption{AraXL \gls{ppa} comparison against SoA laned vector processors.}
    \label{tab:efficiency}
\end{table}

\subsection{SoA comparison}
AraXL is the first \gls{rvv} 1.0 vector processor architecture to feature 64 \glspl{fpu} and a \changes{\acrshort{vlen}} of 64 Kibits (2$\times$ and 4$\times$ compared to the largest count reported so far, respectively). 
AraXL architecture improves Ara2's energy and area efficiencies by 30\% and 50\%, respectively, for the same number of lanes, with a +30\% maximum frequency and no \gls{fpu} utilization drop in the long-vector regime.
Our \todo{32-lane} configuration achieves 4$\times$ the performance of Vitruvius+ with similar area efficiency and the same frequency. Comparing the energy efficiency is harder since the scalar core and cache power consumptions are not included in Vitruvius+'s metric. 

A \gls{ppa} comparison with the VE30 vector engine is not straightforward since only total system power (including interconnects and caches) and area numbers are reported in the literature, making it hard to perform a standalone comparison of the vector unit. Nevertheless, compared to the area evaluations performed in \cite{Minervini2022}, AraXL reaches at least +45\% better area efficiency than the older-generation VE NEC vector unit (10.16 DP-GFLOPS/\(mm^2\) at 1.6 GHz).

%% file: conclusions.tex
In this work, we presented AraXL, a novel RISC-V vector architecture designed to leverage long-vector applications in the \gls{hpc} and \gls{ml} domains. AraXL features the maximum \gls{vrf} size allowed by the V 1.0 specifications (64 Kibit/register) and can scale up to 64 parallel vector lanes with linear scaling ($2\times$ the area with twice the lanes) thanks to dedicated optimizations to the all-to-all interconnects that usually limit the scalability of today's vector processors. 

We implement AraXL in an advanced 22-nm technology node reaching \todo{1.15} GHz and an efficiency of \todo{40.1} GFLOPS/W (0.8V, TT, 25C) for a \todo{64-lane} configuration. AraXL's performance on multiple compute- and memory-intensive kernels doubles when doubling the number of lanes with long vectors in weak-scaling conditions, reaching more than 99\% utilization on sufficiently large matrix multiplications even with 64 lanes.

%% file: ack.tex
\changes{This project was supported in part through the ISOLDE (101112274) project that received funding from the HORIZON CHIPS-JU programme}
\changes{and in part from the Swiss State Secretariat for Education, Research, and Innovation (SERI) under the SwissChips initiative.}